\documentstyle[aps,epsf]{revtex}

\newcommand{\pslash}{p\!\!\!/}

\preprint{UCTP-115-99}

\begin{document}
\draft

\title{On gap equations and color-flavor locking in cold dense
QCD with three massless flavors}

\author{I.A.~Shovkovy$^{a}$\thanks{On leave of absence from Bogolyubov
         Institute for Theoretical Physics, 252143, Kiev,
         Ukraine.}
    and
L.C.R.~Wijewardhana$^{a,b}$}

\address{
$^{a}$Physics Department, University of Cincinnati, 
      Cincinnati, Ohio 45221-0011 \\
$^{b}$Institute of Fundamental Studies, Kandy, Sri Lanka}

\date{October 1, 1999}
\maketitle

\begin{abstract}
The superconductivity in cold dense QCD with three massless flavors
is analyzed in the framework of the Schwinger-Dyson equation. The set 
of two coupled gap equations for the color antitriplet, flavor
antitriplet $(\bar{3},\bar{3})$ and the color sextet, flavor sextet
$(6,6)$ order parameters is derived. It is shown that the
antitriplet-antitriplet gives the dominant contribution to the
color-flavor locked order parameter, while the sextet-sextet is small
but nonzero.  
\end{abstract} 



The idea of color superconductivity in dense QCD is not new
\cite{BarFra,Bail}. However, it is only now that we start to
appreciate the potential relevance of this phenomenon for heavy ion
collisions and for physics of neutron (or quark) stars. The recent
progress was stimulated by the observation that the cold dense quark
matter could undergo a phase transition into a color superconducting
state with the value of the order parameter as large as 100 MeV
\cite{W1,S1}. For comparison, the most optimistic estimates of
Ref.~\cite{Bail} were at the order of 1 MeV or less. Not
surprisingly, the new estimate triggered the burst of studies 
on the subject \cite{other,PR1,Son,Hg,us,SW2,PR2,EHS2,us2}. 

The approach of Refs.~\cite{W1,S1} was based on the analysis of QCD
in the approximation in which the interaction between quarks is
modeled by instantons. Such an approach is best suited for studies of
the low density quark matter. At densities relevant for the
phenomenology of heavy ion collisions and physics of neutron stars,
however, the instanton approach may just start to fail. Therefore, it
was of equally great importance to rederive the same result in the
framework of the weakly coupled QCD at asymptotically large
densities. Along this direction, first it was pointed by Pisarski and
Rischke \cite{PR1} and independently by Son \cite{Son} that the
unscreened gluon modes of magnetic type could provide the interaction
necessary to generate a large value of the order parameter. Later, 
using the approach of the Schwinger-Dyson (SD) equation, it was
shown that this hypothesis was indeed correct \cite{us,SW2,PR2,EHS2}. 

In this paper, we generalize the analysis of
Refs.~\cite{us,SW2,PR2,EHS2} to the case of dense QCD with three
massless flavors. In particular, we pay special attention to the
color-flavor structure of the order parameter. As is known, QCD with
three massless flavors should reveal the color-flavor locked phase
where the original $SU(3)_{C} \times SU(3)_{R} \times SU(3)_{L} 
\times U(1)_{B}$ symmetry of the action breaks down to
$SU(3)_{C+L+R} \times Z_{2}$ \cite{ARW,ABR}. From the symmetry
arguments alone \cite{PR1}, the order parameter should consist of the
color antitriplet, flavor antitriplet  $(\bar{3},\bar{3})$ and
the color sextet, flavor sextet $(6,6)$ contributions. This is also
confirmed by the explicit solution in the effective theory obtained
from the instanton liquid model. There, it is shown that, while the
dominant contribution to the order parameter is the
antitriplet-antitriplet one, the sextet-sextet admixture is still
nonzero, although small \cite{ARW}. Below, we will demonstrate that
the same conclusion remains true when the interaction is mediated by
the perturbative one-gluon exchange.

As in Ref.~\cite{us}, instead of working with Dirac spinors, it is
more convenient to introduce the eight component Majorana spinor,
\begin{equation}
\Psi=\frac{1}{\sqrt{2}} \left(
\begin{array}{c} \psi \\ \psi^{C} \end{array} \right),
\end{equation}
where $\psi^{C}=C\bar{\psi}^{T}$ and $C$ is a unitary charge
conjugation matrix, defined by $C^{-1} \gamma_{\mu}C
=-\gamma_{\mu}^{T}$ and $C=-C^{T}$. In this new notation, the bare
fermion propagator is defined by   
\begin{equation}
\left(G^{(0)}(p)\right)^{-1}=-i\left(\begin{array}{cc} 
\pslash +\mu \gamma^{0} & 0\\ 0 & \pslash -\mu \gamma^{0}
\end{array}\right),
\label{N3-G_0}
\end{equation}
where $\pslash\equiv \gamma^{\nu}p_{\nu}$ and $\mu$ is the chemical
potential. Regarding the notation, we also note that the fermion
field and its propagator bear the color ($a$, $b$, $\dots$) and
flavor ($i$, $j$, $\dots$) indices. Often, however, when it does not
cause any confusion, we will omit them.

The SD equation in the improved rainbow approximation reads \cite{us}
\begin{equation}
\left(G(p)\right)^{-1}_{ab}=
\left(G^{(0)}(p)\right)^{-1}_{ab}
+4\pi\alpha_{s}\int\frac{d^4 q}{(2\pi)^4}
\left(\begin{array}{cc} \gamma^{\mu} & 0   \\
       0  & -\gamma^{\mu}
  \end{array}\right)
\sum_{a^{\prime},b^{\prime}}\sum_{A=1}^{8}
T^{A}_{a^{\prime}a} G_{a^{\prime}b^{\prime}}(q)
T^{A}_{b^{\prime}b}
\left(\begin{array}{cc} \gamma^{\nu} & 0   \\
       0  & -\gamma^{\nu}
  \end{array}\right)
{\cal D}_{\mu\nu}(q-p) .
\label{SD}
\end{equation}
Here ${\cal D}_{\mu\nu}(k)$ is the gluon propagator in the hard
dense loop (HDL) approximation, $\left(G^{(0)}(p)\right)^{-1}_{ab}$
is the inverse of the bare fermion propagator as is given in
Eq.~(\ref{N3-G_0}) above, and $\left(G(p)\right)^{-1}_{ab}$ is the
inverse of the full fermion propagator, defined over the true (color
superconducting) vacuum. The latter, under the assumption of no wave
function renormalization, is given by
\begin{equation}
\left(G(p)\right)^{-1}=-i\left(\begin{array}{cc} 
\pslash +\mu \gamma^{0} & \Sigma \\
\gamma^{0}\Sigma^{\dagger}\gamma^{0} & 
\pslash -\mu \gamma^{0}
\end{array}\right).
\label{N3-G_1}
\end{equation}
The order parameter of the superconducting phase is represented by 
$\Sigma$, appearing in the off diagonal terms.

Without loss of generality, in this paper we consider only the parity
even order parameter. In the instanton approach, it is shown that
such an order parameter determines the true vacuum of the color
superconducting phase \cite{W1,S1}. At sufficiently large density of
quark matter when the instanton density vanishes (and the current
quark masses are small compared to the scale of the condensate), the
parity odd vacuum may have a relatively long life-time and,
therefore, be of great interest too \cite{PR3}. Obviously, the SD
approach with the perturbative kernel, utilized here, is incapable of
differentiating between the parity odd and parity even order
parameters. The generalization of our analysis to the case when both
parity odd and parity even contributions are non-zero is
straightforward. 

Now, let us discuss the color-flavor structure of the order parameter.
In Ref.~\cite{PR1}, from symmetry arguments alone it is argued that
the most general order parameter could contain two pieces: the color
antitriplet, flavor antitriplet, $\Sigma^{ij}_{ab} \sim \delta_{a}^{i}
\delta_{b}^{j} -\delta_{a}^{j} \delta_{b}^{i}$, and the color sextet,
flavor sextet, $\Sigma^{ij}_{ab} \sim \delta_{a}^{i} \delta_{b}^{j}
+\delta_{a}^{j} \delta_{b}^{i}$. Therefore, the most general form of
the solution for the order parameter reads $\Sigma^{ij}_{ab} \sim
\kappa_{1}\delta_{a}^{i} \delta_{b}^{j} +\kappa_{2}\delta_{a}^{j}
\delta_{b}^{i}$ \cite{ARW}. In this paper, we suggest an alternative
(although, equivalent) representation of the color-flavor structure,
\begin{equation}
\Sigma^{ij}_{ab}(p) =\gamma^{5} \sum_{n}
\left[ \Delta^{+}_{n}(p)\Lambda^{(+)}_{p}
+\Delta^{-}_{n}(p)\Lambda^{(-)}_{p}  \right]
\left({\cal P}^{(n)} \right)^{ij}_{ab} 
\label{def-op},
\end{equation}
where 
\begin{equation}
\Lambda^{(\pm)}_{p}=\frac{1}{2}
\left(1\pm \frac{\vec{\alpha}\cdot\vec{p}}{|\vec{p}|}\right)
\label{Lambdas}
\end{equation}
are the free quark (antiquark) on-shell projectors \cite{PR1,SW2},
and ${\cal P}^{(n)}$ are the new projectors in the color-flavor
space, 
\begin{mathletters}
\begin{eqnarray} 
\left({\cal P}^{(1)}\right)_{ab}^{ij}
      &=&\frac{1}{3}\delta_{a}^{i}\delta_{b}^{j}, \\
\left({\cal P}^{(2)}\right)_{ab}^{ij}
      &=&\frac{1}{2}\delta_{ab}\delta^{ij}
      -\frac{1}{2}\delta_{a}^{j}\delta_{b}^{i},  \\
\left({\cal P}^{(3)}\right)_{ab}^{ij}
      &=&\frac{1}{2}\delta_{ab}\delta^{ij}
      +\frac{1}{2}\delta_{a}^{j}\delta_{b}^{i}
      -\frac{1}{3}\delta_{a}^{i}\delta_{b}^{j}.
\end{eqnarray} 
\label{P-i}
\end{mathletters}
These three projectors satisfy the condition of completeness:
$\sum_{n} {\cal P}^{(n)} ={\cal I}$ where ${\cal I}_{ab}^{ij} 
\equiv \delta_{ab}\delta^{ij}$. Also, all of them are symmetric
under the simultaneous exchange of the color and flavor indices
and, as a consequence of this property, $\left( {\cal P}^{(n)}
\right)^{\dagger}={\cal P}^{(n)}$ for $n=1,2,3$.

As we mentioned above, the most general order parameter contains
the antitriplet-antitriplet $(\bar{3},\bar{3})$ and the
sextet-sextet $(6,6)$ contributions. In addition to these, the
three-parameter space swept by our projectors ${\cal P}^{(n)}$
contains an extra piece, the singlet-singlet $\Sigma^{ij}_{ab}
\sim \delta_{ab} \delta^{ij}$. In order to get rid of this singlet,
it is sufficient to introduce the restriction: $\Delta_{3}
=-\Delta_{2}$. Then, as is easy to check, the two independent
parameters, $\Delta^{\pm}_{1}$ and $\Delta^{\pm}_{2}$, are the linear
combinations of the $(\bar{3},\bar{3})$ order parameter
$\Delta^{\pm}_{(\bar{3},\bar{3})}$ and the $(6,6)$ order parameter
$\Delta^{\pm}_{(6,6)}$,
\begin{eqnarray}
\Delta^{\pm}_{1}&=&2\left(\Delta^{\pm}_{(\bar{3},\bar{3})}
+2\Delta^{\pm}_{(6,6)}\right), \label{ant-sym-1}\\
\Delta^{\pm}_{2}&=&\Delta^{\pm}_{(\bar{3},\bar{3})}
-\Delta^{\pm}_{(6,6)}.
\label{ant-sym-2}
\end{eqnarray} 
(Note that our $\Delta^{-}_{1}$ and $\Delta^{-}_{2}$ are related
to $\kappa_1$ and $\kappa_2$ parameters of Ref.~\cite{ARW} as
follows: $\Delta^{-}_{1}=3\kappa_1+\kappa_2$ and $\Delta^{-}_{2}
=-\kappa_2$.) As we will see below, the advantage of working with the
projectors ${\cal P}^{(n)}$ is tremendous, while the price that one
pays for such a convenience is minimal.

By inverting the expression in Eq.~(\ref{N3-G_1}),  we obtain the
propagator,  
\begin{eqnarray} 
G(p) &=&i \left(\begin{array}{cc}
R_{11}(p) & R_{12}(p)\\
R_{21}(p) & R_{22}(p) \end{array}\right),
\label{N3-G}
\end{eqnarray}
where
\begin{mathletters}
\begin{eqnarray}
R_{11}(p)&=&\gamma^{0} \sum_{n} {\cal P}^{(n)} \left(
\frac{p_0+\epsilon_{p}^{-}}{p_0^2-(\epsilon_{p}^{-})^2
-|\Delta^{-}_{n}|^2 }  \Lambda^{(-)}_{p}
+\frac{p_0-\epsilon_{p}^{+}}{p_0^2-(\epsilon_{p}^{+})^2
-|\Delta^{+}_{n}|^2 } \Lambda^{(+)}_{p} 
\right) ,\\
R_{12}(p)&=&\gamma^{5} \sum_{n} {\cal P}^{(n)} \left(
\frac{ \Delta^{+}_{n}(p) }{p_0^2-(\epsilon_{p}^{+})^2
-|\Delta^{+}_{n}|^2 } \Lambda^{(-)}_{p} 
+\frac{\Delta^{-}_{n}(p) }{p_0^2-(\epsilon_{p}^{-})^2
-|\Delta^{-}_{n}|^2 }  \Lambda^{(+)}_{p}
\right) ,\\
R_{21}(p)&=&-\gamma^{5} \sum_{n} {\cal P}^{(n)} \left(
\frac{\left(\Delta^{-}_{n}(p)\right)^{*}}{p_0^2-(\epsilon_{p}^{-})^2
-|\Delta^{-}_{n}|^2 }  \Lambda^{(-)}_{p}
+\frac{\left(\Delta^{+}_{n}(p)\right)^{*}}{p_0^2-(\epsilon_{p}^{+})^2
-|\Delta^{+}_{n}|^2 } \Lambda^{(+)}_{p} 
\right) ,\\
R_{22}(p)&=&\gamma^{0} \sum_{n} {\cal P}^{(n)} \left(
\frac{p_0+\epsilon_{p}^{+}}{p_0^2-(\epsilon_{p}^{+})^2
-|\Delta^{+}_{n}|^2 }  \Lambda^{(-)}_{p}
+\frac{p_0-\epsilon_{p}^{-}}{p_0^2-(\epsilon_{p}^{-})^2
-|\Delta^{-}_{n}|^2 } \Lambda^{(+)}_{p} 
\right),
\end{eqnarray}
\label{Rs}
\end{mathletters}
and $\epsilon_{p}^{\pm}=|\vec{p}|\pm \mu$. In the derivation, we used
the formulae (\ref{a-1}) -- (\ref{a-4}) from Appendix~\ref{appA}.

Now, returning back to the SD equation (\ref{SD}), we see that
the sum over the adjoint indices in the right hand leads to
the following result:
\begin{equation}
\sum_{a^{\prime},b^{\prime}}\sum_{A=1}^{8}
T^{A}_{a^{\prime}a} G^{ij}_{a^{\prime}b^{\prime}}(q)
T^{A}_{b^{\prime}b}=\frac{1}{2}G^{ij}_{ba}(q)-
\frac{1}{6}G^{ij}_{ab}(q).
\end{equation}
At this point, we recall that the color-flavor structure of the quark
propagator is completely determined by that of the projectors in 
Eq.~(\ref{P-i}). Therefore, by using their definition, we derive the
following relations:  
\begin{equation}
\frac{1}{2}\left({\cal P}^{(n)}\right)^{ij}_{ba}-
\frac{1}{6}\left({\cal P}^{(n)}\right)^{ij}_{ab}
=-\frac{1}{3}\sum_{m=1}^{3}A_{mn}
\left({\cal P}^{(m)}\right)^{ij}_{ab},
\label{col-tr}
\end{equation}
where the explicit form of the $3\times 3$ matrix A is given by 
\begin{equation}
A=\left(\begin{array}{rrr}
  0   &   3/2   &  -5/2 \\
 1/2  &  -1/4   &  -5/4 \\
-1/2  &  -3/4   &   1/4 
\end{array} \right).
\end{equation}

By making use of the explicit form of the quark propagator in
Eqs.~(\ref{N3-G}), (\ref{Rs}) and the relations (\ref{col-tr}), 
we derive the gap equation,
\begin{equation}
\Delta^{\pm}_{m}(p)=\frac{2\pi\alpha_{s}}{3}
\sum_{n=1}^{3}A_{mn}\int\frac{d^4 q}{(2\pi)^4}
\left(
\frac{ \Delta^{+}_{n}(q) 
\mbox{~tr}\left[ \Lambda^{(\pm)}_{p}\gamma^{\mu}
\Lambda^{(-)}_{q}\gamma^{\nu}\right]
}{q_0^2-(\epsilon_{q}^{+})^2
-|\Delta^{+}_{n}|^2 } 
+\frac{\Delta^{-}_{n}(q) 
\mbox{~tr}\left[\Lambda^{(\pm)}_{p}\gamma^{\mu}
\Lambda^{(+)}_{q}\gamma^{\nu}\right]
}{q_0^2-(\epsilon_{q}^{-})^2
-|\Delta^{-}_{n}|^2 }  
\right) {\cal D}_{\mu\nu}(q-p) ,
\label{gap-gen}
\end{equation}
which, after taking into account the restriction $\Delta^{\pm}_{3}
=-\Delta^{\pm}_{2}$ discussed earlier, splits into the following set 
of coupled equations:
\begin{eqnarray}
\Delta^{\pm}_{1}(p)&=&\frac{8\pi\alpha_{s}}{3}
\int\frac{d^4 q}{(2\pi)^4} \left( \frac{\Delta^{+}_{2}(q) 
\mbox{~tr}\left[ \Lambda^{(\pm)}_{p}\gamma^{\mu}
\Lambda^{(-)}_{q}\gamma^{\nu}\right]}
{q_0^2-(\epsilon_{q}^{+})^2-|\Delta^{+}_{2}|^2 } 
+\frac{\Delta^{-}_{2}(q) 
\mbox{~tr}\left[\Lambda^{(\pm)}_{p}\gamma^{\mu}
\Lambda^{(+)}_{q}\gamma^{\nu}\right]}
{q_0^2-(\epsilon_{q}^{-})^2-|\Delta^{-}_{2}|^2 }  
\right) {\cal D}_{\mu\nu}(q-p) , \label{eq-D1}\\
\Delta^{\pm}_{2}(p)&=&\frac{\pi\alpha_{s}}{3}
\int\frac{d^4 q}{(2\pi)^4} \left( \frac{\Delta^{+}_{1}(q) 
\mbox{~tr}\left[ \Lambda^{(\pm)}_{p}\gamma^{\mu}
\Lambda^{(-)}_{q}\gamma^{\nu}\right]}
{q_0^2-(\epsilon_{q}^{+})^2 -|\Delta^{+}_{1}|^2 } 
+\frac{\Delta^{-}_{1}(q) 
\mbox{~tr}\left[\Lambda^{(\pm)}_{p}\gamma^{\mu}
\Lambda^{(+)}_{q}\gamma^{\nu}\right]}
{q_0^2-(\epsilon_{q}^{-})^2-|\Delta^{-}_{1}|^2 }  
\right) {\cal D}_{\mu\nu}(q-p) \nonumber\\
&+&\frac{2\pi\alpha_{s}}{3}
\int\frac{d^4 q}{(2\pi)^4} \left( \frac{\Delta^{+}_{2}(q) 
\mbox{~tr}\left[ \Lambda^{(\pm)}_{p}\gamma^{\mu}
\Lambda^{(-)}_{q}\gamma^{\nu}\right]}
{q_0^2-(\epsilon_{q}^{+})^2-|\Delta^{+}_{2}|^2 } 
+\frac{\Delta^{-}_{2}(q) 
\mbox{~tr}\left[\Lambda^{(\pm)}_{p}\gamma^{\mu}
\Lambda^{(+)}_{q}\gamma^{\nu}\right]}
{q_0^2-(\epsilon_{q}^{-})^2-|\Delta^{-}_{2}|^2 }  
\right) {\cal D}_{\mu\nu}(q-p) .\label{eq-D2}
\end{eqnarray}

So far, we did not say much about the form of the gluon propagator 
${\cal D}_{\mu\nu}(q-p)$ used in the kernel of the SD equation.
Apparently, the medium of high density quark matter modifies the
bare gluon propagator in many different ways. The most important 
modification is given by the screening effects
\cite{us,SW2,PR2,EHS2}. By taking these into account in the HDL
approximation, we arrive at the following expression for the gluon
propagator \cite{us}:
\begin{equation}
{\cal D}_{\mu\nu}(ik_4,k_{\perp})\simeq
i \frac{|\vec{k}|}{|\vec{k}|^3+\pi M^2 |k_4|/2} O^{(1)}_{\mu\nu} 
+i \frac{1}{|\vec{k}|^2+|k_4|^2+2M^2}O^{(2)}_{\mu\nu}
+i \frac{\xi}{|\vec{k}|^2+|k_4|^2} O^{(3)}_{\mu\nu},
\label{D-long} 
\end{equation}
where $k=q-p$, $M^2=N_{f} \alpha_{s} \mu^2/\pi$ and the projectors
$O^{(i)}_{\mu\nu}$ are given by \cite{us}
\begin{mathletters}
\begin{eqnarray}
O^{(1)}_{\mu\nu}&=&g^{\perp}_{\mu\nu}
+\frac{k^{\perp}_{\mu}k^{\perp}_{\nu}}{|\vec{k}|^2},
\\
O^{(2)}_{\mu\nu}&=&-\frac{k_0^2}{|\vec{k}|^2} \left(
\frac{k_{\mu}k_{\nu}}{k^2}
-\frac{k_{\mu}\delta^{0}_{\nu}+\delta^{0}_{\mu}k_{\nu}}{k_0}
+\frac{\delta^{0}_{\mu}\delta^{0}_{\nu}}{k_0^2}k^2 \right), \\
O^{(3)}_{\mu\nu}&=& \frac{k_{\mu}k_{\nu}}{k^2}.
\end{eqnarray}
\label{Os}
\end{mathletters}
These projectors define the three types of gluon modes:
$a^{(m)}_{\mu}=O^{1}_{\mu\nu}A^{\nu}$ (magnetic),
$a^{(e)}_{\mu}=O^{2}_{\mu\nu}A^{\nu}$ (electric) and
$a^{(\parallel)}_{\mu}=O^{3}_{\mu\nu}A^{\nu}$ (longitudinal),
respectively. The corresponding three terms in the right hand side
of Eq.~(\ref{D-long}) are the propagators of these three types
of modes. 

Obviously, the gluon propagator is also modified by the Meissner
effect. However, as in the case of the $N_f=2$ QCD, this effect is
negligible everywhere outside the narrow infrared region given by $0
\alt p^2 \alt (M^2 |\Delta_0|)^{2/3}$ \cite{us,SW2,PR2,EHS2}. By
noting that the latter does not overlap with the most important
dynamical region of color symmetry breaking, $|\Delta_0|^2  \alt p^2
\alt \mu^2$, it is justified to neglect the Meissner effect while
working at the leading order in the coupling constant. At the  next to
leading order, however, it will be crucial for the self-consistency
of the solution to properly take this effect into account.

Now, let us return back to the analysis of the gap equations in
Eqs.~(\ref{eq-D1}) and (\ref{eq-D2}). We note that the main
contribution in the right hand side of these equations comes from
the terms proportional to $\Delta_{i}^{-}$ (we assume that $\mu$ is
positive). Then, in our approximate analytical approach, it is
justified to omit the $\Delta_{i}^{+}$ terms altogether. As a
result, the equations for $\Delta^{-}_{i}$ decouple,
\begin{eqnarray}
\Delta^{-}_{1}(p)&=&\frac{8\pi\alpha_{s}}{3}
\int\frac{d^4 q}{(2\pi)^4} \frac{\Delta^{-}_{2}(q) 
\mbox{~tr}\left[\Lambda^{(-)}_{p}\gamma^{\mu}
\Lambda^{(+)}_{q}\gamma^{\nu}\right]}
{q_0^2-(\epsilon_{q}^{-})^2-|\Delta^{-}_{2}|^2 } 
{\cal D}_{\mu\nu}(q-p) , \label{D1}\\
\Delta^{-}_{2}(p)&=&\frac{\pi\alpha_{s}}{3}
\int\frac{d^4 q}{(2\pi)^4} \left(
\frac{\Delta^{-}_{1}(q) 
\mbox{~tr}\left[\Lambda^{(-)}_{p}\gamma^{\mu}
\Lambda^{(+)}_{q}\gamma^{\nu}\right]}
{q_0^2-(\epsilon_{q}^{-})^2 -|\Delta^{-}_{1}|^2 } 
+\frac{2\Delta^{-}_{2}(q) 
\mbox{~tr}\left[\Lambda^{(-)}_{p}\gamma^{\mu}
\Lambda^{(+)}_{q}\gamma^{\nu}\right]}
{q_0^2-(\epsilon_{q}^{-})^2 -|\Delta^{-}_{2}|^2 } 
\right) {\cal D}_{\mu\nu}(q-p) . \label{D2}
\end{eqnarray}
Here we would like to emphasize that only $\Delta^{-}_{i}$ determine
the values of the gaps in the spectrum of quasiparticles at the Fermi
surface. In contrast, $\Delta^{+}_{i}$ just slightly modify the
spectrum of anti-quasiparticles [see Eqs.~(\ref{Rs})].

By following the same arguments as in Ref.~\cite{EHS2}, it can be
shown that the dependence of the solutions for $\Delta^{-}_{i}(q)$ on
the spatial momentum orientation is negligible. Then, the integral
over the angular coordinates, parameterizing the relative orientation
of $\vec{q}$ and $\vec{p}$, can be calculated explicitly in the right
hand side of the gap equations (\ref{D1}) and (\ref{D2}). The results
of these integrations for all types of gluon modes are presented
separately in Eqs.~(\ref{a-13}) --  (\ref{a-18}) in
Appendix~\ref{appA}. 

Next, after neglecting the dependence of $\Delta^{-}_{i}$ on 
$|\vec{q}|$ \cite{us,SW2,PR2,EHS2} and approximating the kernels of
the equations by their dominant contributions at the Fermi surface
[see Eq.~(\ref{a-19}) in Appendix~\ref{appA}], we can also perform
the integration over the absolute value of the spatial momentum. The
result is presented in Eq.~(\ref{a-20}) in Appendix~\ref{appA}. By
using it, we finally arrive at the following system of equations:
\begin{eqnarray}
\Delta^{-}_{1}(p_4) & \simeq & \frac{\nu^2}{4}
\int \frac{d q_4 \Delta^{-}_{2}(q_4)}
{\sqrt{q_4^2+|\Delta^{-}_{2}|^2}} \ln\frac{\Lambda}{|q_4-p_4|},
\label{gap-a-1}\\
\Delta^{-}_{2}(p_4) & \simeq & \frac{\nu^2}{32}
\int \frac{d q_4 \Delta^{-}_{1}(q_4)}
{\sqrt{q_4^2+|\Delta^{-}_{1}|^2}} \ln\frac{\Lambda}{|q_4-p_4|}
+\frac{\nu^2}{16} \int \frac{d q_4 \Delta^{-}_{2}(q_4)}
{\sqrt{q_4^2+|\Delta^{-}_{2}|^2}} \ln\frac{\Lambda}{|q_4-p_4|},
\label{gap-a-2}
\end{eqnarray}
where $\nu=\sqrt{8\alpha_{s}/9\pi}$ and $\Lambda \equiv
16(2\pi)^{3/2}\mu/(3\alpha_{s})^{5/2}$. This set of coupled
equations is solved approximately in Appendix~\ref{appB}. 
Here we present the final result,
\begin{equation}
\Delta^{-}_{1}(p_{4})\approx 2\Delta^{-}_{2}(p_{4})
=-2\Delta^{-}_{3}(p_{4})\approx
2\Delta^{-}_{(\bar{3},\bar{3})}(p_{4}),
\end{equation}
where 
\begin{mathletters}
\begin{eqnarray}
\Delta^{-}_{(\bar{3},\bar{3})}(p_{4}) &\approx &\Delta_{0},
\quad p_{4}\leq \Delta_{0},\\
\Delta^{-}_{(\bar{3},\bar{3})}(p_{4})&\approx&\Delta_{0} 
\sin\left(\frac{\nu}{2}
\ln\frac{\Lambda}{p_{4}}\right),  
\quad  p_{4}\geq \Delta_{0},
\end{eqnarray}
\label{sigma_1}
\end{mathletters}
and
\begin{equation}
\Delta_{0}\approx \Lambda\exp\left(-\frac{\pi}{\nu}\right)
=\frac{16(2\pi)^{3/2}\mu}{(3\alpha_{s})^{5/2}}
\exp\left(-\frac{3\pi^{3/2}}{2^{3/2}\sqrt{\alpha_{s}}}\right).
\end{equation}

By recalling the definition in Eqs.~(\ref{def-op}) and (\ref{P-i}),
we arrive at the following color-flavor structure of the order
parameter: 
\begin{equation} 
\Sigma^{ij}_{ab}(p_{4}) 
\approx\gamma^{5}\left(\delta^{i}_{a}\delta^{j}_{b}
-\delta^{j}_{a}\delta^{i}_{b} \right) 
\Delta^{-}_{(\bar{3},\bar{3})}(p_{4}) \equiv 
\gamma^{5} \sum_{I=1}^{3}\varepsilon^{ijI} \varepsilon_{abI}
\Delta^{-}_{(\bar{3},\bar{3})}(p_{4}) .
\label{op-res}
\end{equation}
This is the dominant antitriplet-antitriplet portion of the solution
for the order parameter in the color-flavor locked phase of QCD with
three massless flavors. However, as one can see from the structure of
the coupled equations (\ref{eq-D1}) and (\ref{eq-D2}), or even from
their approximate version in Eqs.~(\ref{gap-a-1}) and
(\ref{gap-a-2}), the sextet-sextet portion of the order parameter,
$\Delta^{-}_{(6,6)} =(\Delta^{-}_{1}-2\Delta^{-}_{2})/6$, should
also be nonzero. 

Our analysis in Appendix~\ref{appB} [see Eq.~(\ref{b-7}) there]
clearly demonstrates that the repulsion rather than attraction
prevails in the channel with the sextet-sextet order parameter in the
most important dynamical region $\Delta_{0}^2 \alt p^2 \alt \mu^2$.
Therefore, a nonzero sextet-sextet contribution is generated only in
the narrow infrared region $0 \alt p^2 \alt \Delta_{0}^2$ where the
nonlinear character of the gap equations is essential.  Outside this
dynamical region, $\Delta^{-}_{(6,6)}(p)\approx 0$ \footnote{It is
likely, however, that $\Delta^{-}_{(6,6)}(p)\simeq \Delta_0$ for small
momenta, $0 \alt p^2 \alt \Delta_{0}^2$.}. Clearly, a more rigorous
estimate of the magnitude of this sextet-sextet order parameter would
be unreliable in the approximation we use. Indeed, in the infrared
region of momenta, $0 \alt p^2 \alt \Delta_{0}^2$, all the subleading
effects (the Meissner effect, the wave function renormalizations, the
vertex corrections, the corrections from the $\Delta^{+}_{i}$ terms),
which we left aside, should be taken into account more carefully.

Therefore, in further studies, it would be of great interest to
analyze the gap equations up to the next to leading order. 
In addition to giving a proper estimate for the sextet-sextet
contribution, the next to leading corrections could also produce a
finite correction to the magnitude of the antitriplet-antitriplet
order parameter. Indeed, the naive analysis of our gap equations,
that ignores all the subleading corrections but the effect of the
gauge fixing term in the gluon propagator, would modify the value of
the gap as follows: $\Delta^{-}_{(\bar{3},\bar{3})}\to \exp(3\xi/2)
\Delta^{-}_{(\bar{3},\bar{3})}$ [see Eq.~(\ref{a-19})]. Apparently,
the proper consideration of all the subleading effects would result in
cancellation of any such a dependence on $\xi$. It is difficult,
however, to show this explicitly, and it is not clear what kind of a
correction this may produce for the value of the gap. 

In conclusion, in this paper we derived the coupled set of gap
equations for the order parameters of the color-flavor locked phase
of massless $N_f=3$ dense QCD. Our analysis of these equations shows
that the dominant contribution to the order parameter comes in the
color antitriplet, flavor antitriplet channel. The other, the color
sextet, flavor sextet contribution is small but non-zero. 

After this paper was completed, we learned about the paper 
by T.~Schafer discussing somewhat similar issues \cite{Schafer}.

\begin{acknowledgments}
L.C.R.W. thanks Professor K.~Tennakone for his hospitality at the
Institute of Fundamental Studies. This work is supported by U.S.
Department of Energy  Grant No. DE-FG02-84ER40153.   
\end{acknowledgments}

\appendix

\section{Formulae used in derivation of gap equations}
\label{appA}

In derivation of Eqs.~(\ref{Rs}), we find it convenient to use the
following representations for the kinetic terms of the spinor
propagators:
\begin{eqnarray}
\pslash+\mu\gamma^{0} &=& \gamma^{0}\left[
(p_0-\epsilon^{-}_{p}) \Lambda^{(+)}_{p}
+(p_0+\epsilon^{+}_{p}) \Lambda^{(-)}_{p} \right], \label{a-1}\\
\pslash-\mu\gamma^{0} &=& \gamma^{0}\left[
(p_0-\epsilon^{+}_{p}) \Lambda^{(+)}_{p}
+(p_0+\epsilon^{-}_{p}) \Lambda^{(-)}_{p} \right], \label{a-2}\\
\left(\pslash+\mu\gamma^{0}\right)^{-1} &=& \gamma^{0}\left[
\frac{1}{p_0+\epsilon^{+}_{p}} \Lambda^{(+)}_{p}
+\frac{1}{p_0-\epsilon^{-}_{p}} \Lambda^{(-)}_{p} \right], 
\label{a-3}\\
\left(\pslash-\mu\gamma^{0}\right)^{-1} &=& \gamma^{0}\left[
\frac{1}{p_0+\epsilon^{-}_{p}} \Lambda^{(+)}_{p}
+\frac{1}{p_0-\epsilon^{+}_{p}} \Lambda^{(-)}_{p} \right], 
\label{a-4}
\end{eqnarray}
where the projectors $\Lambda^{(\pm)}_{p}$ are defined in
Eq.~(\ref{Lambdas}) in the main text.

In order to perform the angular integration in the right hand sides 
of the gap equations (\ref{eq-D1}) and (\ref{eq-D2}), one first need
to calculate the following two types of traces over the Dirac
indices:
\begin{eqnarray}
\mbox{~tr}\left[\Lambda^{(\pm)}_{p}\gamma^{\mu}
\Lambda^{(\pm)}_{q}\gamma^{\nu}\right] &=& 
g^{\mu\nu} (1+t )-2 g^{\mu 0}g^{\nu 0} t 
+\frac{\vec{q}^{\mu}\vec{p}^{\nu}+\vec{q}^{\nu}\vec{p}^{\mu}}
{|\vec{q}| |\vec{p}|}+\dots, \label{a-5}\\
\mbox{~tr}\left[\Lambda^{(\pm)}_{p}\gamma^{\mu}
\Lambda^{(\mp)}_{q}\gamma^{\nu}\right] &=& 
g^{\mu\nu} (1-t )+2 g^{\mu 0}g^{\nu 0} t 
-\frac{\vec{q}^{\mu}\vec{p}^{\nu}+\vec{q}^{\nu}\vec{p}^{\mu}}
{|\vec{q}| |\vec{p}|}+\dots,
\label{a-6}
\end{eqnarray}
where $t =\cos\theta$ is the cosine of the angle between
three-vectors $\vec{q}$ and $\vec{p}$, and irrelevant antisymmetric
terms are denoted by ellipsis.

By contracting these traces with the projectors of the magnetic,
electric and longitudinal types of gluon modes, defined in
Eqs.~(\ref{Os}), we arrive at  
\begin{mathletters}
\begin{eqnarray}
O^{(1)}_{\mu\nu} \mbox{~tr}\left[\Lambda^{(\pm)}_{p}\gamma^{\mu}
\Lambda^{(\pm)}_{q}\gamma^{\nu}\right] &=& 2 (1+t )
\frac{q^2+p^2-q p(1+t )}
{q^2+p^2-2q pt }, \label{a-7}\\
O^{(1)}_{\mu\nu} \mbox{~tr}\left[\Lambda^{(\pm)}_{p}\gamma^{\mu}
\Lambda^{(\mp)}_{q}\gamma^{\nu}\right] &=&2 (1-t )
\frac{q^2+p^2+q p(1-t )}
{q^2+p^2-2q pt }, \label{a-8}\\
O^{(2)}_{\mu\nu} \mbox{~tr}\left[\Lambda^{(\pm)}_{p}\gamma^{\mu}
\Lambda^{(\pm)}_{q}\gamma^{\nu}\right] &=&2 (1-t )
\frac{q^2+p^2+q p(1-t )}{q^2+p^2-2q pt }
- (1-t )\frac{(q+p)^2+(q_4-p_4)^2} {q^2+p^2-2q pt 
+(q_4-p_4)^2}, \label{a-9}\\
O^{(2)}_{\mu\nu} \mbox{~tr}\left[\Lambda^{(\pm)}_{p}\gamma^{\mu}
\Lambda^{(\mp)}_{q}\gamma^{\nu}\right] &=&2 (1+t )
\frac{q^2+p^2-q p(1+t )}{q^2+p^2-2q pt }
-(1+t ) \frac{(q-p)^2+(q_4-p_4)^2}{q^2+p^2-2q pt 
+(q_4-p_4)^2}, \label{a-10}\\
O^{(3)}_{\mu\nu} \mbox{~tr}\left[\Lambda^{(\pm)}_{p}\gamma^{\mu}
\Lambda^{(\pm)}_{q}\gamma^{\nu}\right] &=& (1-t )
\frac{(q+p)^2+(q_4-p_4)^2}
{q^2+p^2-2q pt 
+(q_4-p_4)^2}, \label{a-11}\\
O^{(3)}_{\mu\nu} \mbox{~tr}\left[\Lambda^{(\pm)}_{p}\gamma^{\mu}
\Lambda^{(\mp)}_{q}\gamma^{\nu}\right] &=& (1+t )
\frac{(q-p)^2+(q_4-p_4)^2}
{q^2+p^2-2q pt +(q_4-p_4)^2},
\label{a-12}
\end{eqnarray}
\label{Otr}
\end{mathletters}
where $q\equiv |\vec{q}|$, $p\equiv |\vec{p}|$, $q_4\equiv -i q_0$
and $p_4\equiv -i p_0$.

Now, by making use of the explicit expressions (\ref{Otr}), we
easily perform the angular integrations of all types appearing in
the gap equations,
\begin{eqnarray}
&&\int \frac{d\Omega |\vec{q}-\vec{p}|}
{|\vec{q}-\vec{p}|^3+\omega_{l}^3}
O^{(1)}_{\mu\nu} \mbox{~tr}\left[\Lambda^{(\pm)}_{p}\gamma^{\mu}
\Lambda^{(\pm)}_{q}\gamma^{\nu}\right] = \pi \Bigg[
-\frac{2}{qp}+\frac{(q^2-p^2)^2+\omega_{l}^4}
{\sqrt{3}\omega_{l}^2 q^2 p^2}\arctan
\left(\frac{\sqrt{3}\omega_{l}\mbox{min}(q,p)}
{\omega_{l}^2+|q^2-p^2|-\omega_{l}\mbox{max}(q,p)}\right)
 \nonumber\\
&&+\frac{(q^2-p^2)^2-\omega_{l}^4} {3\omega_{l}^2 q^2 p^2}
\ln\frac{\omega_{l}+|q+p|}{\omega_{l}+|q-p|}
-\frac{(q^2-p^2)^2-\omega_{l}^4} {6\omega_{l}^2 q^2 p^2}
\ln\frac{\omega_{l}^2+|q+p|^2-\omega_{l}|q+p|}
{\omega_{l}^2+|q-p|^2-\omega_{l}|q-p|}+\frac{4}{3qp}
\ln\frac{\omega_{l}^3+|q+p|^3}{\omega_{l}^3+|q-p|^3}
\Bigg],
\label{a-13}
\end{eqnarray}
where $\omega_{l}^3=(\pi/2) M^2 \omega$ and $\omega =|q_4-p_4|$. 
Similarly,
\begin{eqnarray}
&&\int \frac{d\Omega |\vec{q}-\vec{p}|}
{|\vec{q}-\vec{p}|^3+\omega_{l}^3}
O^{(1)}_{\mu\nu} \mbox{~tr}\left[\Lambda^{(\pm)}_{p}\gamma^{\mu}
\Lambda^{(\mp)}_{q}\gamma^{\nu}\right] = \pi \Bigg[
\frac{2}{qp}-\frac{(q^2-p^2)^2+\omega_{l}^4}
{\sqrt{3}\omega_{l}^2 q^2 p^2}\arctan
\left(\frac{\sqrt{3}\omega_{l}\mbox{min}(q,p)}
{\omega_{l}^2+|q^2-p^2|-\omega_{l}\mbox{max}(q,p)}\right)
 \nonumber\\
&&-\frac{(q^2-p^2)^2-\omega_{l}^4} {3\omega_{l}^2 q^2 p^2}
\ln\frac{\omega_{l}+|q+p|}{\omega_{l}+|q-p|}
+\frac{(q^2-p^2)^2-\omega_{l}^4} {6\omega_{l}^2 q^2 p^2}
\ln\frac{\omega_{l}^2+|q+p|^2-\omega_{l}|q+p|}
{\omega_{l}^2+|q-p|^2-\omega_{l}|q-p|}+\frac{4}{3qp}
\ln\frac{\omega_{l}^3+|q+p|^3}{\omega_{l}^3+|q-p|^3}
\Bigg],
\label{a-14}
\end{eqnarray}
\begin{eqnarray}
&&\int \frac{ d\Omega}{|\vec{q}-\vec{p}|^2+\omega^2+2M^2}
O^{(2)}_{\mu\nu} \mbox{~tr}\left[\Lambda^{(\pm)}_{p}\gamma^{\mu}
\Lambda^{(\pm)}_{q}\gamma^{\nu}\right] =\frac{2\pi}{q p}
+\frac{\pi}{2q^2 p^2} \Bigg[
-\frac{(q^2-p^2)^2}{2M^2+\omega^2}
\ln\frac{(q+p)^2}{(q-p)^2}\nonumber\\ &&
-\frac{\left[(q-p)^2+2M^2+\omega^2\right]
\left[(2M^2+\omega^2)^2+(q+p)^2\omega^2\right]}
{2M^2(2M^2+\omega^2)}
\ln\frac{(q+p)^2+2M^2+\omega^2}{(q-p)^2+2M^2+\omega^2}\nonumber\\ &&
+\frac{[(q+p)^2+\omega^2][(q-p)^2+\omega^2]}{2M^2}
\ln\frac{(q+p)^2+\omega^2}{(q-p)^2+\omega^2} \Bigg],
\label{a-15}
\end{eqnarray}
\begin{eqnarray}
&&\int \frac{ d\Omega}{|\vec{q}-\vec{p}|^2+\omega^2+2M^2}
O^{(2)}_{\mu\nu} \mbox{~tr}\left[\Lambda^{(\pm)}_{p}\gamma^{\mu}
\Lambda^{(\mp)}_{q}\gamma^{\nu}\right] =-\frac{2\pi}{q p}
+\frac{\pi}{2q^2 p^2} \Bigg[
\frac{(q^2-p^2)^2}{2M^2+\omega^2}
\ln\frac{(q+p)^2}{(q-p)^2}\nonumber\\ &&
+\frac{\left[(q+p)^2+2M^2+\omega^2\right]
\left[(2M^2+\omega^2)^2+(q-p)^2\omega^2\right]}
{2M^2(2M^2+\omega^2)}
\ln\frac{(q+p)^2+2M^2+\omega^2}{(q-p)^2+2M^2+\omega^2}\nonumber\\ 
&&-\frac{[(q+p)^2+\omega^2][(q-p)^2+\omega^2]}{2M^2}
\ln\frac{(q+p)^2+\omega^2}{(q-p)^2+\omega^2} \Bigg],
\label{a-16}
\end{eqnarray}
\begin{eqnarray}
\xi \int \frac{ d\Omega }{|\vec{q}-\vec{p}|^2+\omega^2}
O^{(3)}_{\mu\nu} \mbox{~tr}\left[\Lambda^{(\pm)}_{p}\gamma^{\mu}
\Lambda^{(\pm)}_{q}\gamma^{\nu}\right] &=& \pi \xi\Bigg[
-\frac{2}{qp}+\frac{(q+p)^2+\omega^2}{2q^2 p^2}
\ln\frac{(q+p)^2+\omega^2}{(q-p)^2+\omega^2} \Bigg] ,
\label{a-17} \\
\xi \int \frac{ d\Omega }{|\vec{q}-\vec{p}|^2+\omega^2}
O^{(3)}_{\mu\nu} \mbox{~tr}\left[\Lambda^{(\pm)}_{p}\gamma^{\mu}
\Lambda^{(\mp)}_{q}\gamma^{\nu}\right] &=& \pi \xi\Bigg[
\frac{2}{qp}-\frac{(q-p)^2+\omega^2}{2q^2 p^2}
\ln\frac{(q+p)^2+\omega^2}{(q-p)^2+\omega^2} \Bigg].
\label{a-18}
\end{eqnarray}

As a result, the angular average of the gluon propagator (multiplied
by $q^2$ weight) in the vicinity of the Fermi surface is given by
the following approximate expression:
\begin{eqnarray}
q^2 \int d\Omega 
{\cal D}_{\mu\nu}(q-p) \mbox{~tr}\left[
\Lambda^{(\pm)}_{p}\gamma^{\mu} \Lambda^{(\mp)}_{q}\gamma^{\nu}
\right] &\approx& 2i\pi \left[
\frac{2}{3}\ln\frac{(2\mu)^3}{|\epsilon_{q}^{-}|^3+\omega_{l}^3}
+\ln\frac{(2\mu)^2}{(\epsilon_{q}^{-})^2+2M^2+\omega^2}
+\xi\right] ,
\label{a-19}
\end{eqnarray}
where $\epsilon_{q}^{-}=q-\mu$ and $\xi$ if the gauge fixing parameter.
The three terms in the last expression are the leading contributions
of the magnetic, electric and the longitudinal gluon modes,
respectively. Because of the absence of the logarithmic factor in
front of the gauge fixing parameter, the longitudinal gluon modes
become relevant only in the next to leading order. Therefore, here 
we ignore it in what follows. (Notice, however, that the solution
to the gap equation might still be very sensitive to the choice of
the gauge in the approximation with no wave function renormalization
taken into account.) 
Finally, the integral over the spatial momentum gives
\begin{eqnarray}
&&\int_{0}^{\mu}\frac{q^2 dq }
{q_4^2+(\epsilon_{q}^{-})^2+|\Delta|^2} 
\int d\Omega  {\cal D}_{\mu\nu}(q-p) \mbox{~tr}\left[
\Lambda^{(\pm)}_{p}\gamma^{\mu} \Lambda^{(\mp)}_{q}\gamma^{\nu}
\right] \nonumber\\
&\approx& \frac{4i\pi^2}{3\sqrt{q_4^2+|\Delta|^2}}
\left(\ln\frac{(2\mu)^3}{\omega_{l}^3}
+\frac{3}{2}\ln\frac{(2\mu)^2}{2M^2+\omega^2}\right)
\approx \frac{4i\pi^2}{3\sqrt{q_4^2+|\Delta|^2}}
\ln\frac{\Lambda}{\omega} ,
\label{a-20}
\end{eqnarray}
where $\Lambda\equiv (2\mu)^6/(\sqrt{2}\pi M^5)
=16(2\pi)^{3/2}\mu/(N_{f}\alpha_{s})^{5/2}$.

\section{The approximate solution of the gap equations}
\label{appB}

In order to solve the set of integral equations (\ref{gap-a-1})
and (\ref{gap-a-2}), we further approximate the kernels and introduce
finite infrared and ultraviolet cutoffs,
\begin{eqnarray}
\Delta_{1}(p) & \simeq & \frac{\nu^2}{2}\left(
 \int_{\Delta_{0}}^{p} \frac{d q}{q} \Delta_{2}(q)
 \ln\frac{\Lambda}{p}
+\int_{p}^{\Lambda} \frac{d q}{q} \Delta_{2}(q)
 \ln\frac{\Lambda}{q} \right),
\label{b-1}\\
\Delta_{2}(p) & \simeq & \frac{\nu^2}{16}\left(
 \int_{\Delta_{0}}^{p} \frac{d q}{q} \Delta_{1}(q)
\ln\frac{\Lambda}{p}
+\int_{p}^{\Lambda} \frac{d q}{q}\Delta_{1}(q)
 \ln\frac{\Lambda}{q} \right)
+\frac{\nu^2}{8} \left( \int_{\Delta_{0}}^{p} \frac{d q}{q}
\Delta_{2}(q) \ln\frac{\Lambda}{p}
+\int_{p}^{\Lambda} \frac{d q}{q}\Delta_{2}(q)
 \ln\frac{\Lambda}{q} \right),
\label{b-2}
\end{eqnarray}
where $\nu=\sqrt{8\alpha_{s}/9\pi}$ and $\Lambda \equiv
16(2\pi)^{3/2}\mu/(3\alpha_{s})^{5/2}$. Also, for brevity of
notation, we omit the superscript ``-" at $\Delta_{i}$ in this
appendix. The set of equations (\ref{b-1}) and (\ref{b-2}) is
equivalent to the following two coupled differential equations:
\begin{eqnarray}
&& p \Delta_{1}^{\prime\prime}(p)+\Delta_{1}^{\prime}(p)
+\frac{\nu^2}{2} \frac{\Delta_{2}(p)}{p}=0, \label{b-3}\\
&& p \Delta_{2}^{\prime\prime}(p)+\Delta_{2}^{\prime}(p)
+\frac{\nu^2}{16}\frac{\Delta_{1}(p)}{p}
+\frac{\nu^2}{8}\frac{\Delta_{2}(p)}{p}=0, 
\label{b-4}
\end{eqnarray}
along with the following infrared and ultraviolet boundary
conditions:
\begin{equation}
\left. p \Delta_{1,2}^{\prime}(p)\right|_{p=\Delta_{0}}=0
\quad\mbox{(IR)},\qquad
\Delta_{1,2}(\Lambda)=0
\quad\mbox{(UV)}.
\label{b-5}
\end{equation}

By introducing new variables, $\Delta_{(\bar{3},\bar{3})}$ and 
$\Delta_{(6,6)}$, as in Eqs.~(\ref{ant-sym-1}) and (\ref{ant-sym-2}),
the two differential equations decouple,
\begin{eqnarray}
p  \Delta_{(\bar{3},\bar{3})}^{\prime\prime}(p)
  +\Delta_{(\bar{3},\bar{3})}^{\prime}(p) 
  +\frac{\nu^2}{4}\frac{\Delta_{(\bar{3},\bar{3})}(p)}{p}=0, \label{b-6}\\ 
p  \Delta_{(6,6)}^{\prime\prime}(p) 
  +\Delta_{(6,6)}^{\prime}(p)  
  -\frac{\nu^2}{8}\frac{\Delta_{(6,6)}(p)}{p}=0. \label{b-7}
\end{eqnarray}
Both the infrared and ultraviolet boundary conditions for
$\Delta_{(\bar{3},\bar{3})}$ and $\Delta_{(6,6)}$ are the same as
those for $\Delta_{1,2}(p)$, given in Eq.~(\ref{b-5}). The negative
sign in front of the $\nu^2$ term in Eq.~(\ref{b-7}) indicates that
the repulsion, rather than attraction, dominates in the pairing
channel with the $(6,6)$ order parameters $\Delta_{(6,6)}(p)$. 
This explains, in particular, why the only solution for
$\Delta_{(6,6)}(p)$, satisfying the boundary conditions in
Eq.~(\ref{b-5}), is the trivial one. The solution for
$\Delta_{(\bar{3},\bar{3})}(p)$, on the other hand, reads  
\begin{equation} 
\Delta_{(\bar{3},\bar{3})}(p)= \Delta_{0} \sin\left(\frac{\nu}{2}
\ln\frac{\Lambda}{p}\right), \quad 
\Delta_{0} =\Lambda \exp\left(-\frac{\pi}{\nu}\right),
\label{b-8} 
\end{equation}
where, in order to be consistent with the approximation used in
derivation of Eqs.~(\ref{b-1}) and (\ref{b-2}), we defined the
overall constant of the solution by requiring that
$\Delta_{(\bar{3},\bar{3})}(\Delta_{0}) =\Delta_{0}$.

\end{document}